%% file: MAIN.tex
\documentclass[a4paper, oneside, 10pt]{article}
\usepackage{eurosym}
\usepackage{libertine}
\usepackage[english]{babel}
\usepackage[utf8]{inputenc}
\usepackage{enumerate}
\usepackage{graphicx}
\usepackage[hidelinks]{hyperref}
\usepackage{fancyhdr}
\usepackage{amsthm}
\usepackage{amssymb}
\usepackage{amsmath}
\usepackage{amsfonts}
\usepackage{minibox}
\usepackage{float}
\usepackage{graphicx}
\usepackage{subcaption}
\usepackage[export]{adjustbox}
\usepackage{wrapfig}
\usepackage[strict]{changepage}
\usepackage{listings}
\usepackage{listingsutf8}
\usepackage{color}
\usepackage{tikz}
\usetikzlibrary{calc}
\usepackage{pgfplots}
\pgfplotsset{width=13cm,compat=1.9}
\usepgfplotslibrary{external}
\usepackage{framed}
\usepackage{braket}
\usepackage{longtable}
\usepackage{epigraph}
\usepackage[numbers,sort&compress]{natbib}
\usepackage{ragged2e}
\usepackage{caption}

\numberwithin{equation}{section}
\input xypic
\theoremstyle{remark}

\DeclareMathOperator{\M}{\mathcal{M}}

\date{}
\title{\textbf{Cosmological Dynamics of Accelerating Model in $f(T)$ Gravity with Special Forms of Deceleration Parameter}}
\author{Shivank Pal$^{1}$ \thanks{Email: shivankpal1997@gmail.com}, Romanshu Garg$^{2}$\thanks{Email: romanshugarg18@gmail.com}, Gyan Prakash Singh$^{2}$\thanks{Email: gpsingh@mth.vnit.ac.in}, Gauree Shanker$^{1}$\thanks{Email:  gauree.shanker@cup.edu.in}
\vspace{0.3cm}\\
${}^{1}$ Department of Mathematics and Statistics,\\Central University of Punjab, Bathinda, 151401, Punjab, India. 
\vspace{0.3cm}\\
${}^{2}$ Department of Mathematics,\\ Visvesvaraya National Institute of Technology,\\ Nagpur, 440010, Maharashtra, India.
}

\begin{document}
\input{Chapters/ABSTRACT}
\input{Chapters/INTRODUCTION}
\input{Chapters/FTFORMALISM}

\input{Chapters/DECELERATIONPARAMETER}
\input{Chapters/OBSERVATIONALCONSTRAINTS}
\input{Chapters/COSMOLOGICALPARAMETERS}

\input{Chapters/STATEFINDERDIAGNOSTIC}
\input{Chapters/CONCLUSION}

\bibliographystyle{unsrtnat}
\bibliography{REFERENCES}

\end{document}

%% file: Chapters/ABSTRACT.tex
\maketitle
\begin{abstract}
In this paper, the dynamical behavior of the accelerated expansion of the universe is discussed within the framework of $f(T)$ gravity, considering the power law functional form $ f(T) = \alpha (-T)^n.$  Two distinct redshift-dependent parameterizations of the deceleration parameter such as \( q(z) = q_0 + q_1 \left( \frac{\ln(N+z)}{z+1} - \ln N \right) \) and $q(z) = \dfrac{1}{2} + \dfrac{q_1 z + q_2}{(1+z)^2}$ are considered. We have derived the Hubble parameter in terms of redshift and discussed its effect on other cosmological parameters. Using Bayesian statistical analysis and $\chi^2$-minimization,  the median values of the model parameters for both the cosmic chronometer(CC) and the joint (CC + Pantheon) dataset have been determined. Further, energy density, pressure, the equation of state for Dark Energy, Energy conditions and statefinder diagnostics are analyzed. The current age of the universe is also computed for these models.



\end{abstract}
    

%% file: Chapters/INTRODUCTION.tex
\section{Introduction}
It is well established on the basis of a variety of cosmological observations\cite{1,2,2020A&A...641A...6P} that the rate of expansion of the universe has been increasing in current epochs.
A prominent candidate for late-time acceleration is dark energy (DE), whose exact form continues to be mysterious. Cosmological observations estimate that dark energy and dark matter together make up approximately $95-96 \%$ of the universe \cite{41}. It is widely accepted that one of the most promising candidates for dark energy is the dynamical cosmological term $\Lambda$. However, the $\Lambda$-term model faces two important problems, namely the fine-tuning problem and the coincidence problem \cite{copeland06,carroll01,padmanabhan03,di2021realm}. The observational evidences have motivated cosmologists to develop theoretical models describing the universe, indicating its accelerated expansion. Researchers have proposed various approaches, including modifications to Einstein's field equations by considering different forms of matter distribution and proposing alternative gravity theories to explain this cosmic acceleration.
To address current cosmological challenges and to find a more suitable model of the universe, several approaches have been proposed, among which modified gravity theories remain the most widely studied. In this sequence, several modified theories have been widely explored \cite{7,8,nojiri11,nojiri17,harko2011f,elizalde10,bamba10,harko10,capozziello19,capozziello2023,hulke20,kotambkar2017anisotropic,garg24,mandal2023,lalke23,singh2024,garg2025late,singh2022cosmic,singh2022lagrangian,singh1997new,singh2024conservative,goswami2024flrw}.
\par A fascinating modification of gravity that has captivated cosmologists is the so-called \( f(T) \) teleparallel gravity \cite{11,12}. \( f(T) \) gravity analogously extends the idea of \( f(R) \) gravity, where the gravitational field equations are derived from a Lagrangian that is a function of the Ricci scalar\( (R) \) of the underlying geometry. However, instead of employing the torsion-free Levi-Civita connection of General Relativity, it relies on the curvature-free Weitzenböck connection. 
 The $f(T)$ gravity theory has attracted considerable attention, leading to numerous studies exploring diverse cosmological phenomena. For instance, investigations have focused on obtaining solutions that describe the universe's evolution \cite{13} and understanding its thermodynamic behavior \cite{14}.  
 Furthermore, efforts have been made to map the universe's expansion history through cosmography \cite{19}, examine fundamental constraints on the energy content through energy conditions \cite{20}, and explore alternative early universe scenarios like matter bounce cosmologies \cite{21}. Various Observational data have also been used to constrain the parameters of $f(T)$ models \cite{22}. Zhadyranova et al. \cite{23} conducted an in-depth study of late-time cosmic acceleration by analyzing a linear \( f(T) \) cosmological model with observational data. In addition, a comprehensive review of the $f(T)$ gravity framework is available in reference \cite{24}. Bamba et al. studied the cosmological evolutions of the equation of state for dark energy $\omega_{DE}$ in the exponential and logarithmic as well as their combination $f(T)$ theories\cite{15}. Paliathanasis et al. present a complete Noether symmetry analysis in the framework of $f(T)$ gravity\cite{paliathanasis2014new}. Capozziello et al. proposed a model-independent formalism to numerically solve the modified Friedmann equations in the framework of $f(T)$ teleparallel cosmology\cite{capozziello2017model}.  Numerous investigations have been completed in relation to the $f (T)$ gravity theory \cite{zhadyranova2024exploring, duchaniya2024,maurya24,shekh25,maurya22, maurya2023anisotropic, bhar2024anisotropic, bamba2016bounce, kavya2024can, chaudhary2023constraints, duchaniya2022dynamical,nunes2016new,mandal2020temporal,maurya2024role,das2023study,chakraborty2023classical,dixit2021probe,ren2022gaussian}.\\
\par In the present paper, we study the parameterized dark energy with two different special forms of the deceleration parameter in $f(T)$ gravity. 
Specifically, our study is based on the power law form of $f(T)$ expressed as \( f(T) = \alpha (-T)^n \), where \( \alpha \) and \( n \) are arbitrary constants. Model parameters are constrained by utilizing observational datasets, including the cosmic chronometer and joint (CC + Pantheon) datasets. This research work particularly emphasizes a critical assessment of the late-time accelerated expansion in the $f(T)$ gravity model by analyzing various cosmological parameters. \\
\par The structure of this paper is as follows. The section (\ref{sec:2}) presents an outline of the essential mathematical formulation of $f(T)$ gravity and field equations corresponding to the FLRW metric, laying the foundation for exploring cosmological solutions. In section (\ref{sec:3}), two distinct parameterization forms of the deceleration parameter are considered to derive an analytical expression for the Hubble parameter. Section (\ref{sec:4}) is dedicated to constraining the model parameters through a Bayesian statistical analysis, utilizing observational data from cosmic chronometers and the joint(CC+Pantheon) dataset.
The behavior of physical quantities (like pressure, energy density), energy conditions and statefinder diagnostic for these models are discussed in Section (\ref{sec:5}). Additionally, the age of the universe for these models is also estimated. Finally, in section (\ref{sec:6}), a comprehensive conclusion is provided.

%% file: Chapters/FTFORMALISM.tex
\section{Mathematical Framework for $f(T)$ Gravity }\label{sec:2}
Here, we discuss the basic mathematical formalism of $f(T)$ gravity. Based on the torsion scalar, $f(T)$ gravity represents a modified theory whose geometric action is determined by an algebraic function associated with the torsion. The line element is given as
\begin{equation*}
    ds^2 = g_{\mu \nu} dx^{\mu}dx^{\nu} = \eta_{lm} \theta^l \theta^m,
    \tag{1} \label{equ(1)}
\end{equation*}
with the components 
\begin{equation*}
    dx^{\mu} = e_l^{~ \mu} \theta^l, \quad \theta^l = e_{~ \mu}^l dx^{\mu},
\end{equation*}
here, $\eta_{lm}= diag(1, -1,-1,-1)$ is the metric associated with flat spacetime, and $\{e_{~ \mu}^l \}$ are the components of the tetrad. These tetrads satisfy the conditions
\begin{equation*}
    e_l^{~ \mu} e_{~ \nu}^l = \delta_{\nu}^{\mu}, \quad e_{\mu}^{~ ~~l} e_{~~ m}^{\mu} = \delta_m^l.
\end{equation*}
The fundamental connection used in $f(T)$ gravity is the Weitzenböck connection \cite{25}, defined as
\begin{equation*}
    \Gamma_{\mu \nu}^\alpha = e_l^{~ \alpha} \partial_{\mu}e_{~ \nu}^l = - e_{~ \mu}^l \partial_{\nu}e_l^{~\alpha}.
    \tag{2} \label{equ(2)}
\end{equation*}
With this connection, the components of the corresponding torsion tensor are expressed as
\begin{equation*}
    T_{~~\mu \nu}^{\alpha} = \Gamma_{~~ \nu \mu}^{\alpha} - \Gamma_{\mu \nu}^{\alpha} = e_l^{~~ \alpha} \left(\partial_{\mu}e_{~~\nu}^l - \partial_{\nu} e_{\mu}^l \right).
    \tag{3} \label{equ(3)}
\end{equation*}
Based on the torsion tensor, the contorsion tensor is given as
\begin{equation*}
    K_{~\alpha}^{\mu \nu} = - \frac{1}{2} \left(T_{~\alpha}^{\mu \nu} - T_{\alpha}^{\nu \mu} - T_{\alpha}^{~\mu \nu} \right),
    \tag{4} \label{equ(4)}
\end{equation*}
which, when combined with the torsion tensor, yields the tensor
\begin{equation*}
    S_{\alpha}^{~ \mu \nu} = \frac{1}{2} \left(K_{~\alpha}^{\mu \nu} + \delta_{\alpha}^{\mu} T_{~\lambda}^{\lambda \mu} - \delta_{\alpha}^{\nu} T_{~ \lambda}^{\lambda \mu} \right).
    \tag{5} \label{equ(5)}
\end{equation*}
The torsion scalar, denoted as $T$, is a quantity constructed from the torsion tensor and $S_{\alpha}^{~\mu \nu}$, is given as \cite{12,26}
\begin{equation*}
    T = S_{\alpha}^{~ \mu \nu}T_{~\mu \nu}^{\alpha} = \frac{1}{2} T^{\alpha \mu \nu}T_{\alpha \mu \nu} + \frac{1}{2} T^{\alpha \mu \nu}T_{\nu \mu \alpha} - T_{\alpha \mu}^{~ \alpha} T_{~ \nu}^{\nu \mu}.
    \tag{6} \label{equ(6)}
\end{equation*}
Further, the action for this gravity is expressed as \cite{11, 44}
\begin{equation*}
S = \frac{1}{2 \kappa^2} \int d^4x e[T + f(T)] + \int d^4 x e L_m,
\tag{7} \label{equ(7)}
\end{equation*}
where \( e \) represents the determinant of the tetrad, expressed as \( e = \text{det}(e^l_{~\mu}) = \sqrt{-g} \).
By varying the action \eqref{equ(7)} with respect to the tetrads, the field equations of $f(T)$ gravity are obtained as
\begin{equation*}
    S_{\mu}^{~ \nu \rho} \partial_{\rho}T f_{TT} + \left[e^{-1} e^l_{\mu} \partial_{\rho} \left(ee_l^{~\mu} S_{\alpha}^{~ \nu \lambda} \right) + T_{~ \lambda \mu}^{\alpha}S_{\alpha}^{~\nu \lambda} \right]f_T + \frac{1}{4} \delta_{\mu}^{\nu} f = \frac{\kappa^2}{2}T_{\mu}^{\nu},
    \tag{8} \label{equ(8)}
\end{equation*}
where $f_T = \frac{\partial f}{\partial T}, f_{TT} = \frac{\partial^2 f}{\partial T^2}$, and $T_{~ \mu}^{\nu}$ represents the energy-momentum tensor, given by
\begin{equation*}
    T_{~\mu}^{\nu} = (\rho + p)u_{\mu}u^{\nu} - p \delta_{\mu}^{\nu},
    \tag{9} \label{equ(9)}
\end{equation*}
 where $p$ and $\rho$ represent pressure and energy density, respectively, of the ordinary matter constituents of the universe. The four-velocity field associated with this ordinary matter, $u^{\mu}$, satisfies the condition $u^{\mu}u_{\nu} =1$.
 \vspace{0.2cm}\\
In this study, we consider the flat FLRW metric, which is commonly adopted to facilitate the application of the aforementioned theory in a cosmological context. This choice leads to the derivation of the modified Friedmann equations. The flat FLRW spacetime metric is mathematically expressed as follows
\begin{equation*}
    ds^2 = dt^2 - a^2(t)\delta_{lm}dx^{l}dx^{m}, 
    \tag{10} \label{equ(10)}
\end{equation*}
where $a(t)$ is the scalar factor. Consequently, the torsion scalar can be evaluated for the line element \eqref{equ(10)} as $T = -6H^2$.\\
The Friedmann equations corresponding to the metric \eqref{equ(10)} are \cite{12}
\begin{equation*}
    6H^2 + 12 H^2 f_T + f = 2 \kappa^2 \rho,
    \tag{11} \label{equ(11)}
\end{equation*}
\begin{equation*}
    2\left(2 \dot H + 3H^2 \right) + f +4 \left( \dot H + 3H^2 \right)f_T - 48 H^2 \dot H f_{TT} = -2 \kappa^2 p.
    \tag{12} \label{equ(12)}
\end{equation*}
Here, the ``dot" notation denotes differentiation with respect to cosmic time \( t \), while \( H \) represents the Hubble parameter. Furthermore, \( \rho \) and \( p \) denote the energy density and pressure of the matter content, respectively. Assuming $\kappa^2 = 1$, we can express the equations \eqref{equ(11)} and \eqref{equ(12)} as
\begin{equation*}
    3H^2 = \rho + \rho_{DE},
    \tag{13} \label{equ(13)}
\end{equation*}
\begin{equation*}
    -2 \dot H - 3 H^2 = p + p_{DE}.
    \tag{14} \label{equ(14)}
\end{equation*}
Here, the energy density and pressure attributed to DE are established as
\begin{equation*}
    \rho_{DE} = -6H^2 f_T - \frac{1}{2}f,
    \tag{15} \label{equ(15)}
\end{equation*}
\begin{equation*}
    p_{DE} = \frac{1}{2}f + 2 (\dot H + 3H^2)f_T + 2 H \dot f_T.
    \tag{16} \label{equ(16)}
\end{equation*}
From equations \eqref{equ(15)} and \eqref{equ(16)}, we obtain the expression of the EoS parameter of DE as,
\begin{equation*}
    \omega_{DE} = \frac{p_{DE}}{\rho_{DE}} = -1 - \frac{4(H \dot f_T + \dot H f_T)}{12 H^2 f_T + f}.
    \tag{17} \label{equ(17)}
\end{equation*}

%% file: Chapters/DECELERATIONPARAMETER.tex
\section{Deceleration Parameter Parameterization}\label{sec:3}
\par The deceleration parameter $q$ plays a pivotal role in describing the expansion of the universe. It characterizes the nature of cosmic expansion: a negative value $(q <0)$ signifies accelerated expansion, while a positive value $(q>0)$ indicates deceleration. When the value of $q$ is below $-1$, the universe undergoes super-accelerated expansion. Specific values of $q$ correspond to distinct cosmological phases: $q= -1$ represents the de-Sitter phase, $q = \frac{1}{2}$ corresponds to the matter-dominated era, and $q=1$ describes the radiation-dominated era.
In this regard, previous studies have adopted various parametric forms of $q$, while some researchers have focused on nonparametric forms. These approaches have been widely examined in the literature to characterize fundamental cosmological challenges, including the initial singularity problem, the problem of all time decelerating expansion, the horizon problem, and Hubble tension, among other key issues\cite{banerjee05,cunha08,escamilla22}. Various parametrization of the deceleration parameter $q(z)$ have been extensively studied\cite{mandal2023,cunha08,gadbail2022parametrization,shekh2024late,arora2024diagnostic}.
\par With this motivation,
the present study considers two distinct functional forms of deceleration parameter in terms of the redshift $z$ as  
$q(z) = q_0 + q_1 \left( \frac{\ln(N +z)}{z+1} - \ln N \right)$ \cite{mamon2017parametric} and $q(z) = \dfrac{1}{2} + \dfrac{q_1 z + q_2}{(1+z)^2}$ \cite{gong2006observational} to describe the dymnamics of the universe in $f(T)$ gravity.
Here $q_0$, $q_1$, $q_2$, and $N$ are some dimensionless parameters. 
The time derivative of the Hubble parameter satisfies the relation $\dot{H}=-(1+q(z))H^2$. Consequently, an integral relation can be derived that gives the relation between the Hubble parameter and deceleration parameter
\begin{equation*}
    H(z) = H_0 \exp \left[\int_0^z (1 + q(z)) d \ln(1+z) \right].
    \tag{19} \label{equ(19)}
\end{equation*}
\textbf{Model-1} We adopt the parametrization of the deceleration parameter given as\cite{mamon2017parametric}
\begin{equation*}
    q(z) = q_0 + q_1 \left( \frac{\ln(N +z)}{z+1} - \ln N \right)
    \tag{20} \label{equ(20)}
\end{equation*}
By employing the equations $\eqref{equ(19)}$ and $\eqref{equ(20)}$, we get the Hubble parameter in terms of redshift $z$ as
\begin{equation*}
    H(z) = H_0 (z+1)^{1 + q_0 - q_1 \ln N + \frac{q_1}{N-1}} (z+N)^{- \left(\frac{q_1}{z+1} + \frac{q_1}{N-1} \right)} (N)^{\frac{Nq_1}{N-1}},
    \tag{21} \label{equ(21)}
\end{equation*}
here, \( H_0 \) denotes the Hubble constant.
\vspace{0.3cm} \\[3pt]
\noindent \textbf{Model-2} Another adopted parameterization of the deceleration parameter is given as\cite{gong2006observational}:
\begin{equation*}
    q(z) = \dfrac{1}{2} + \dfrac{q_1 z + q_2}{(1+z)^2}
    \tag{22} \label{equ(22)}
\end{equation*}
Using equations $\eqref{equ(19)}$ and $\eqref{equ(22)}$, we get the Hubble parameter in terms of the redshift $z$ as:
\begin{equation*}
    H(z) = H_0 (1+z)^{\frac{3}{2}} \exp \left[\dfrac{(q_1 + q_2)z^2 + 2 q_2z}{2(1+z)^2} \right]
     \tag{23} \label{equ(23)}
\end{equation*}
\begin{figure}[h]
    \centering
    \includegraphics[width=11.0cm, height=8.0cm]{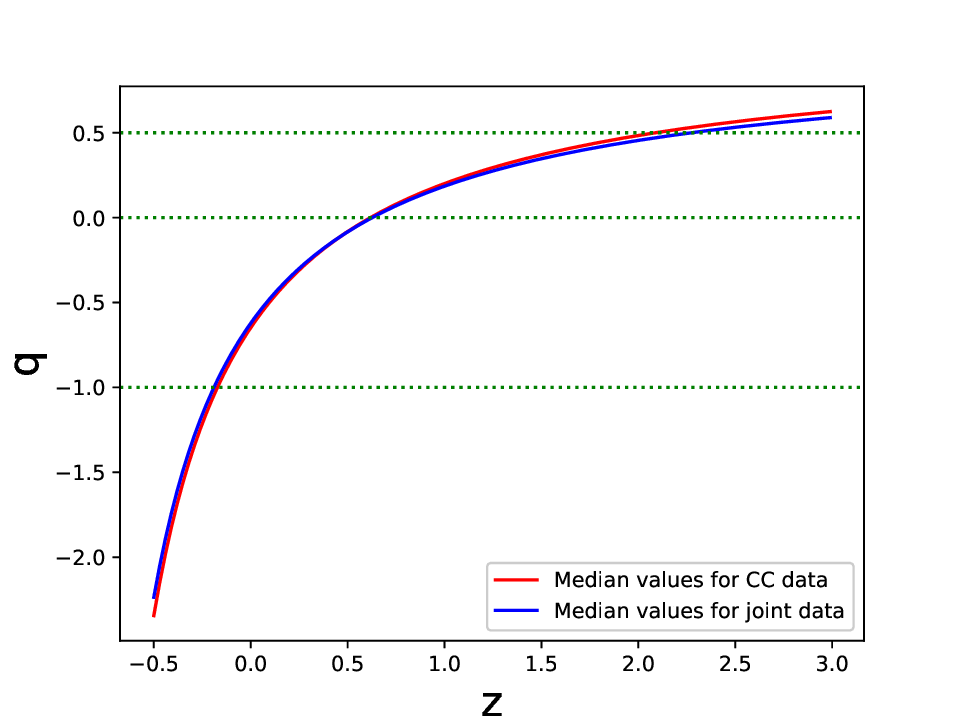}
    \caption{{\bf{For Model-1:}} $q(z)$ with $z$}
    \label{fig: Figure 3}
\end{figure}

\begin{figure}[h]
    \centering
   \includegraphics[width=11.0cm, height=8.0cm]{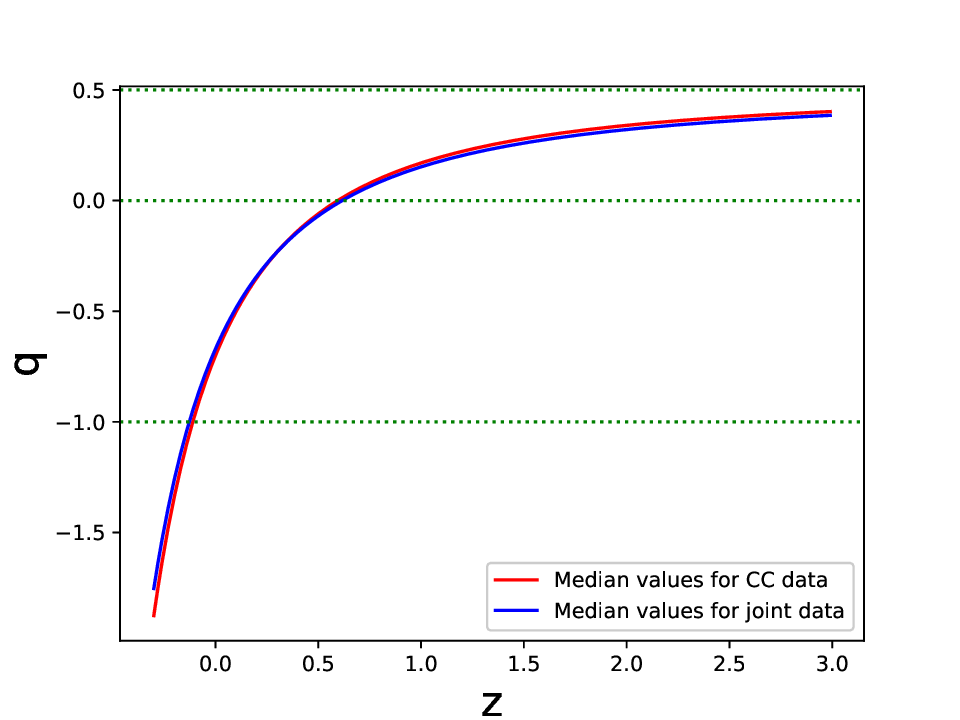}
    \caption{{\bf{For Model-2:}} $q(z)$ with $z$}
    \label{fig: Figure 8}
\end{figure}

\par The analysis of the deceleration parameter for the CC and the joint data sets is presented in figures (\ref{fig: Figure 3}) and (\ref{fig: Figure 8}). Figures (\ref{fig: Figure 3}) and (\ref{fig: Figure 8}) demonstrate the transition of the universe's evolution from a decelerated to an accelerated expansion phase. For median values of model parameters, the current values of the deceleration parameter for model-1 are $q_0 = -0.648$, $q_0 = -0.622 $ for the CC data and the joint data, respectively, and for model-2 are $q_0 = -0.7$(CC) and $-0.67$(joint). The negative values indicate that the universe is currently undergoing an accelerated expansion at $(z=0)$.  It has been demonstrated that the deceleration parameter supports the present phase of accelerated expansion. For model-1, this transition occurs at $z = 0.62$ and $z = 0.63$ for the median values for the CC data and the joint data set, respectively. Similarly, for model-2, the value of transition redshift is $z=0.6$(for CC data) and $z=0.62$(for joint data). 
\par  For both data estimates, the deceleration parameter $q= \frac{1}{2}$, highlights a matter-dominated past in the both models. The universe might experience super-exponential acceleration in the future due to the influence of phantom-like dark energy. Since the accelerated expansion of the universe crosses the de-Sitter line, we can term this form of dark energy as quintom dark energy, as mentioned in the reference \cite{zhao2006quintom}.


%% file: Chapters/OBSERVATIONALCONSTRAINTS.tex
\section{Observational Constriants}\label{sec:4}
This section analyzes the compliance of the parameterized Hubble parameters ($\eqref{equ(21)}$, $\eqref{equ(23)}$) with the cosmic chronometer(CC) and joint( CC+Pantheon) dataset. To conduct the statistical analysis, we employ $\chi^2$-minimization and Bayesian techniques, utilizing the Markov Chain Monte Carlo (MCMC) method through the emcee Python library \cite{46}. This approach helps constrain the model parameters\textquotesingle ~median values. 
\subsection{The Cosmic Chronometer Data}\label{sec:4.1}
In order to constrain model parameters, we use the cosmic chronometer dataset. Using a dataset of $31$ cosmic chronometer data points \cite{27,28} derived from the differential ages (DA) of galaxies, spanning a red-shift range of  $0.007 \leq z \leq 1.965$, we determine the median values of the model parameters. The foundational principle of cosmic chronometer observations, introduced by JImenez and Loeb\cite{49}, established a relationship between the Hubble parameter$(H)$, cosmic time$(t)$ and red-shift$(z)$ expressed as $H(z) = \frac{-1}{1+z} \frac{dz}{dt}$. We determine the model parameters through the minimization of the $\chi^2$ function (which is comparable to the maximizing likelihood function) expressed as\cite{50,51,52}. 
{\begin{equation*}
    \chi^2_{CC} (\theta) = \sum_{i=1}^{31} \dfrac{[H_{th}(\theta, z_i) - H_{obs}]^2}{\sigma^2_{H(z_i)}},
    \tag{24} \label{equ(24)}
\end{equation*}
where $H_{th}(\theta, z_i)$ denotes the theoretical values of the Hubble parameter, while $H_{obs}(z_i)$ represents the observed values of the Hubble parameter. The term $\sigma_{H(z_i)}^2$ corresponds to the standard deviation from observations. 
\par Figure (\ref{fig: Figure 1}) illustrates the error bars for the CC data points along with the best fit Hubble parameter curve.  

\begin{figure}[h]
    \centering
    \includegraphics[width=1.0\linewidth]{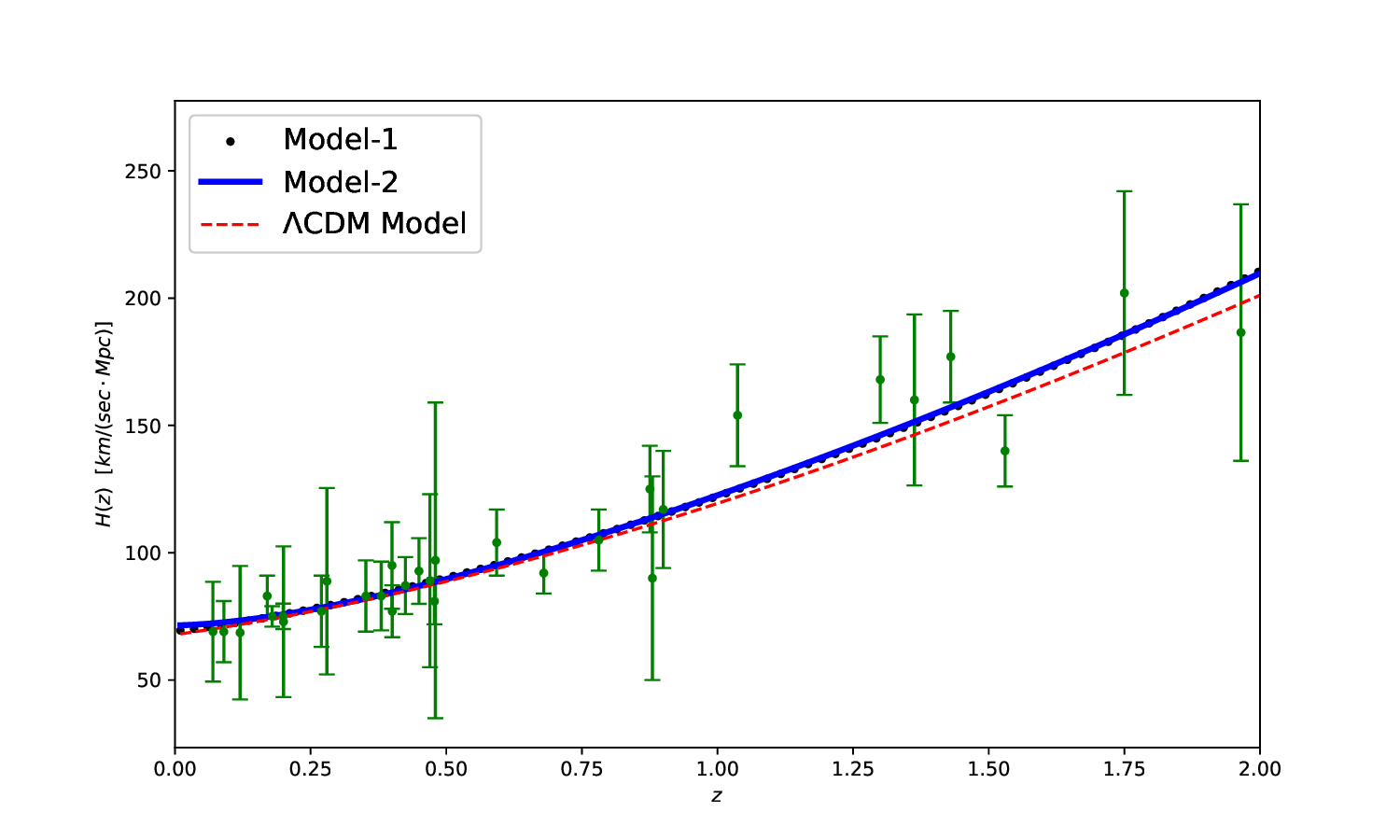}
    \caption{The best-fit $H(z)$ curve for the present models is compared with the $\Lambda$CDM model.}
    \label{fig: Figure 1}
\end{figure}

\subsection{The Pantheon Data}\label{sec:4.2}
The Pantheon sample, comprising 1048 Type Ia supernova (SNIa) data points within the redshift range \(0.01 < z < 2.26\)\cite{31}, is utilized in this study. The SNIa dataset is assembled from multiple surveys, including CfA1–CfA4 \cite{29,30}, the Pan-STARRS1 Medium Deep Survey \cite{31}, SDSS \cite{32}, SNLS\cite{33} and the Carnegie Supernova Project (CSP) \cite{34}. For the analysis based on Pantheon data, the apparent magnitude \( \mu_{th}(z) \) is expressed as  
\begin{equation*}
    \mu_{th}(z) = M + 5 \log_{10} \left[\dfrac{d_L (z)}{M pc} \right] +25,
    \tag{25} \label{equ(25)}
\end{equation*}  
with \( M \) being the absolute magnitude. \( d_L(z) \) represents luminosity distance (which has the dimension of length) and is defined in the following manner \cite{35} 
\begin{equation*}
    d_L(z) = c (1+z) \int_0^z \dfrac{dz'}{H(z')},
    \tag{26} \label{equ(26)}
\end{equation*}  
here \( z \) signifies the redshift of SNIa as ascertained in the cosmic microwave background (CMB) rest frame, and \( c \) is the speed of light. The luminosity distance \( d_L \) is frequently reformulated in its dimensionless Hubble-free luminosity form as \( D_L(z) = H_0 d_L(z)/c \). Accordingly, equation \eqref{equ(25)} can be rewritten as follows
\begin{equation*}
    \mu_{th}(z) = M + 5 \log_{10} \left[\dfrac{c/H_0}{Mpc} \right] +25 + 5 \log_{10} [D_L(z)].
    \tag{27} \label{equ(27)}
\end{equation*}
There will be a degeneracy between $H_0$ and $M$ in the $\Lambda$CDM model framework\cite{ellis2012relativistic,asvesta2022observational}. We express \( \mathcal{M} \) as a combination of these parameters as shown below.
\begin{equation*}
    \M = 25 + 5 \log_{10} \left[\dfrac{c/H_0}{Mpc} \right] +M = M +42.38 - 5 \log_{10}(h),
    \tag{28} \label{equ(28)}
\end{equation*}
here, \( H_0 = h \times 100 \) km/(sec·Mpc). This parameter is incorporated into the relevant \( \chi^2 \) expression for the Pantheon data during the MCMC analysis as \cite{36,50,51,52,53}
\begin{equation*}
    \chi_P^2 = \nabla \mu_i C_{ij}^{-1} \nabla \mu_j,
    \tag{29} \label{equ(29)}
\end{equation*}
here, \( \nabla \mu_i = \mu_{obs}(z_i) - \mu_{th}(z_i) \), \( C_{ij}^{-1} \) denotes the inverse of the covariance matrix, and \( \mu_{th} \) is given by the equation \eqref{equ(27)}. Since the luminosity distance depends on the Hubble parameter, we employ the emcee package\cite{37} along with the relevant equation to obtain the maximum likelihood estimate using the joint (CC+Pantheon) dataset. The joint $\chi^{2}$ for the maximum likelihood estimate is defined as \( \chi_{CC}^2 + \chi_P^2 \). Figures (\ref{fig: Figure 2}) and (\ref{fig: Figure 9}) are exhibited for 1$\sigma$ and 2$\sigma$ likelihood contour maps and $1D$ distributions with MCMC analysis through the combination of CC+Pantheon datasets.
The resulting median values of model parameters from the MCMC analysis are presented in Table (\ref{tab: Table 1}) and Table (\ref{tab: Table 2}).
\begin{table}[H]
\centering
\scalebox{0.71}{\begin{tabular}{|c|c|c|c|c|c|c|c|c|}
\hline
\textbf{Dataset} & \textbf{$H_0 [Km/(sec.Mpc)]$} & \textbf{$q_0$} & \textbf{$q_1$} & \textbf{$N$} & \textbf{$\M$} & \textbf{$z_t$} & \textbf{$\omega_0$} & \textbf{$t_0$ (Gyr)}   \\ [1.0ex]\hline 
\textbf{CC} & $69.21_{-0.78}^{+0.78}$ & $-0.648_{-0.061}^{+0.061}$ & $-0.557_{-0.058}^{+0.069}$ & $22^{+6}_{-7}$ & - & $0.62$ & $-0.6857$ & $13.07$   \\ [1.5ex] \hline
\textbf{CC} + Pantheon & $68.9_{-1.9}^{+1.9}$ & $-0.622_{-0.078}^{+0.078}$ & $-0.53_{-0.11}^{+0.11}$ & $22^{+3}_{-3}$ & $23.804_{-0.014}^{+0.014}$ & $0.63$ &  $-0.6722$ &  $13.19$ \\ [1.5ex]\hline
\end{tabular}}
\caption{{\bf{For Model-1:}} For both CC and joint datasets, the median values of the model parameters and the present values of $q_0$, $w_0$, and $t_0$.}
\label{tab: Table 1}
\end{table}

\begin{table}[H]
\centering
\scalebox{0.74}{\begin{tabular}{|c|c|c|c|c|c|c|c|c|}
\hline
\textbf{Dataset} & \textbf{$H_0 [Km/(sec.Mpc)]$} & \textbf{$q_1$} & \textbf{$q_2$} & \textbf{$\M$} & \textbf{$q_0$} & \textbf{$z_t$} & \textbf{$\omega_0$}&  \textbf{$t_0$ (Gyr)}   \\ [1.0ex] \hline
\textbf{CC} & $69.3_{-4.8}^{+6.6}$ & $-0.12_{-0.71}^{+1.2}$ & $-1.20^{+0.25}_{-0.76}$ & - & $-0.7$ & $0.6$ & $-0.8$ & $13.67$    \\ [1.5ex]\hline
\textbf{CC} + Pantheon & $69.1_{-1.9}^{+1.9}$ & $-0.22_{-0.43}^{+0.43}$ & $-1.17^{+0.10}_{-0.10}$ & $23.801_{-0.015}^{+0.015}$ & $-0.67$ & $0.62$ & $-0.78$  &  $13.78$  \\ [1.5ex]\hline
\end{tabular}}
\caption{{\bf{For Model-2:}} For both CC and joint datasets, the median values of the model parameters and the present values of $q_0$, $w_0$, and $t_0$.}
\label{tab: Table 2}
\end{table}

\begin{figure}
	\begin{center}
\includegraphics[width=17.5cm, height=15.5cm]{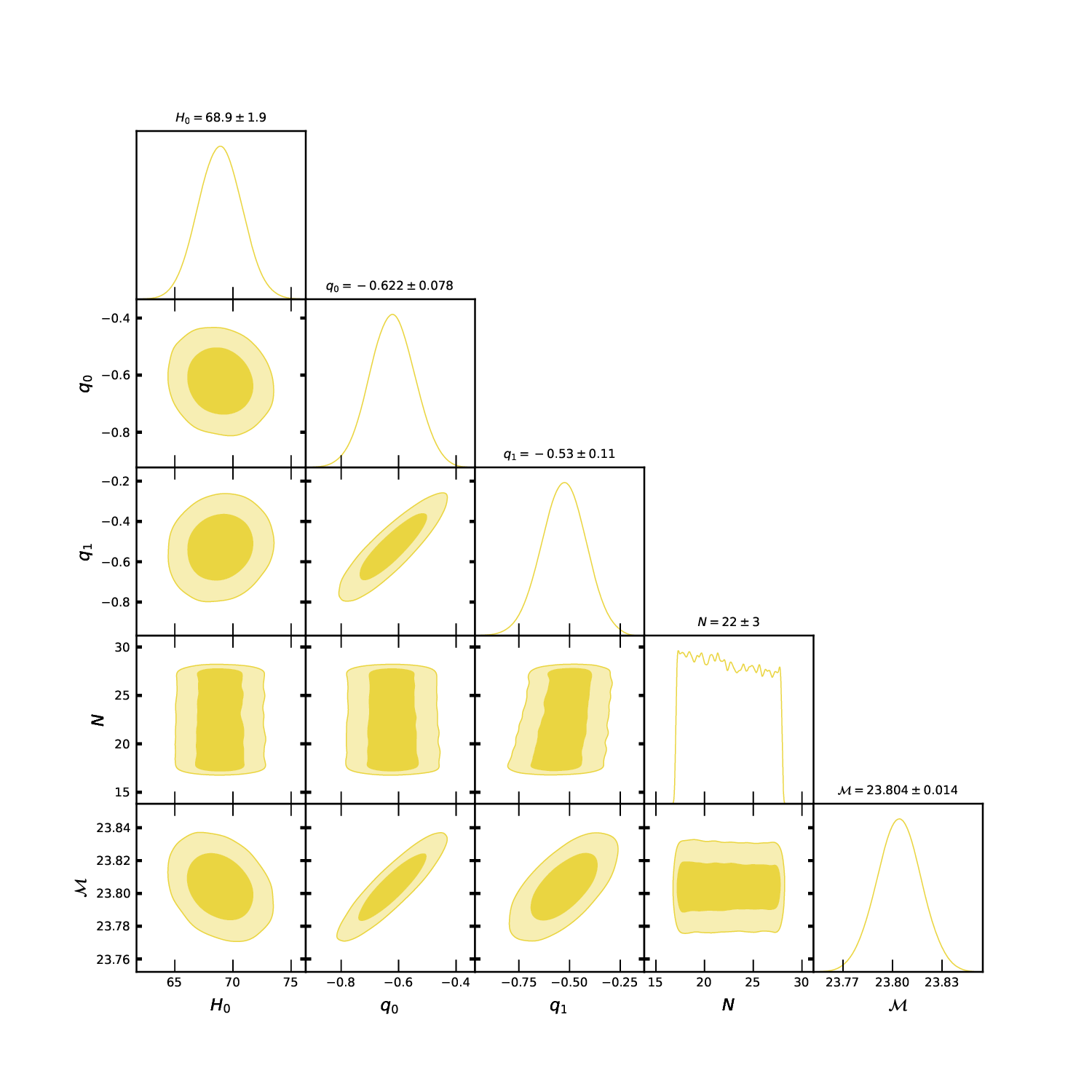}
\caption{{\bf{For Model-1:}} Marginalized 1D and 2D posterior contour map with median values of $H_0$, $q_0$, $q_1$ and $N$ using Joint data set.}
\label{fig: Figure 2}
\end{center}
\end{figure}

\begin{figure}[h]
    \centering
    \includegraphics[width=1.0\linewidth]{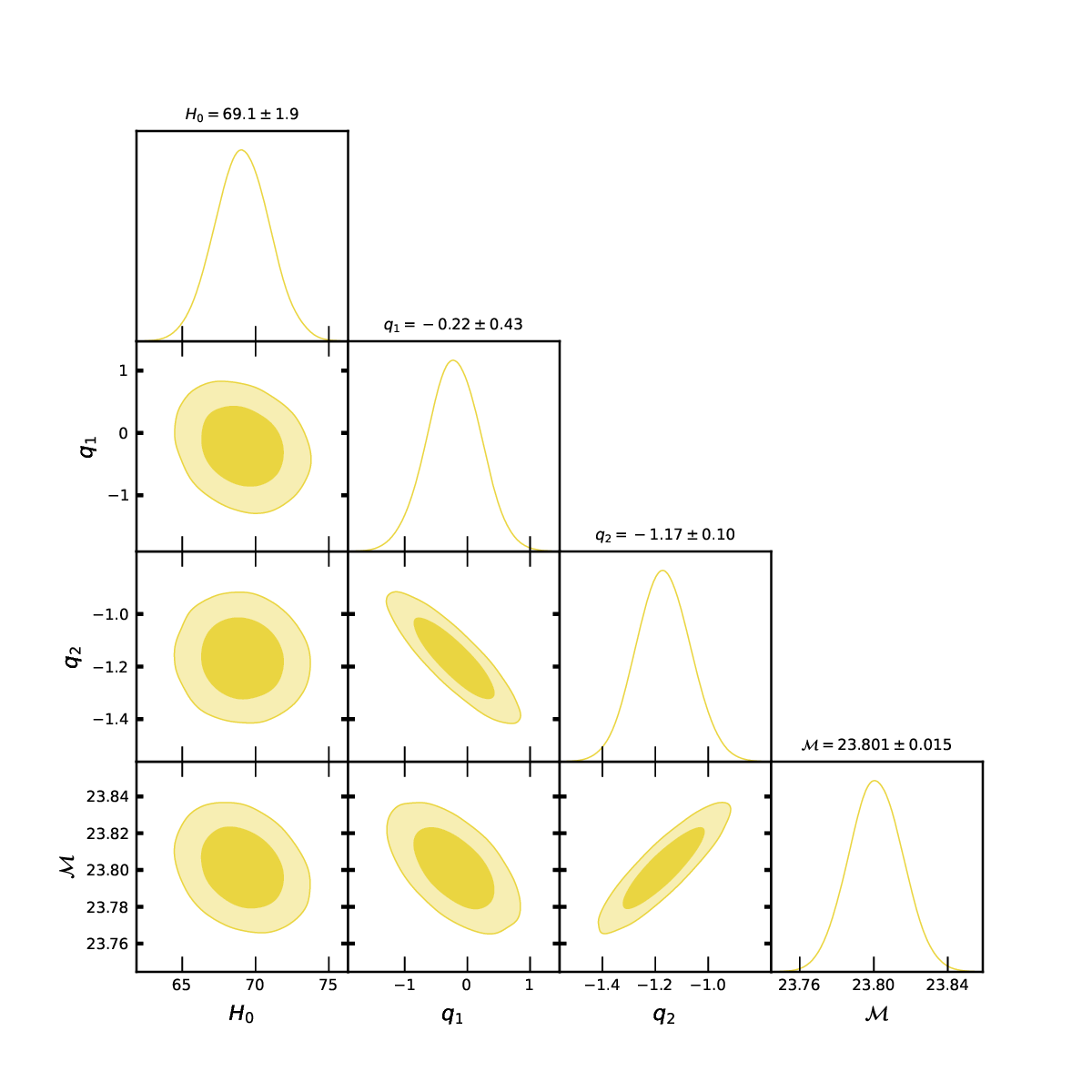}
    \caption{{\bf{For Model-2:}} Marginalized 1D and 2D posterior contour map with median values of $H_0$, $q_1$ and $q_2$ using Joint data set.}
    \label{fig: Figure 9}
\end{figure}



%% file: Chapters/COSMOLOGICALPARAMETERS.tex
\section{Cosmic Dynamics and the Physical Behavior  }\label{sec:5}

\subsection{The Physical Behavior of Models}
We examine the physical properties exhibited by fundamental quantities such as energy density and pressure. For the constrained parameters, the energy density remains positive during all expansion history, while the pressure appears to have transitioned to negative value in the recent cosmological past. As the universe evolves from decelerated to accelerated expansion the energy density maintains its positive nature, while due to the dark energy dominance the pressure becomes negative.
\vspace{0.3cm} \\ 
By utilizing the \eqref{equ(15)}, \eqref{equ(16)}, and \eqref{equ(21)}, we can express the energy density for DE $(\rho_{DE})$ and pressure for DE $(p_{DE})$, for model-1 as
\begin{equation*}
        \rho_{DE} = \alpha \left(n - \frac{1}{2}\right) 6^n \left[H_0 (z +1)^a (z+N)^{-b} N^ {\left(\frac{Nq_1}{N-1}\right)}  \right]^{2n}, \ \ \ \text{(for\  Model-1)}
        \tag{30} \label{equ(30)}
\end{equation*}
\scriptsize
\begin{equation*}
    p_{DE}(z) = \alpha(1-2n)6^{n-1}\left[H_0 (z +1)^a (z+N)^{-b} N^ {\left(\frac{Nq_1}{N-1}\right)}  \right]^{2n}  \left[3 - 2n(z +1) \left(\frac{a}{z+1} - \frac{b}{z+N} + \frac{q_1}{(z+1)^2} \ln (z+N) \right) \right],\ \ \ \text{(for\  Model-1)}
    \tag{31} \label{equ(31)}
\end{equation*}
\normalsize
\vspace{0.2cm}\\
where $a = 1 + q_0 -q_1 \ln N + \frac{q_1}{N-1}$ and $b = \frac{q_1}{z+1} + \frac{q_1}{N-1}$.
\vspace{0.6cm}\\
By using the \eqref{equ(15)}, \eqref{equ(16)} and \eqref{equ(23)}, we can express the energy density for DE $(\rho_{DE})$ and pressure for DE $(p_{DE})$, for model-2 as
\vspace{0.6cm}\\
\begin{equation*}
    \rho_{DE} = \alpha \left(n- \frac{1}{2} \right)6^n \left[H_0 (1+z)^{\frac{3}{2}} \exp \left[\dfrac{(q_1 + q_2)z^2 + 2 q_2z}{2(1+z)^2} \right] \right]^{2n} \ \ \ \text{(for\  Model-2)}
    \tag{32} \label{equ(32)}
\end{equation*}

\begin{equation}
\begin{split}
    p_{DE} = & -\alpha 6^{2n-1} \Bigg[ \,3 H_0^{2n}(1+z)^{3n} 
    \exp\!\left( 2n \, \frac{(q_1 + q_2)z^2 + 2 q_2 z}{2(1+z)^2} \right) \\
    & + \; 2n H_0^{2n+1}(1+z)^{3n + \tfrac{5}{2}}
    \exp\!\left( (2n+1)\,\frac{(q_1 + q_2)z^2 + 2 q_2 z}{2(1+z)^2} \right) \\
    \qquad &\left(\tfrac{3}{2} + \frac{zq_1 + q_2}{(1+z)^2} \right) 
    \Bigg]
\end{split} \ \ \ \text{(for\  Model-2)}
\tag{33}\label{equ(33)}
\end{equation}

For the constrained values of the model parameters, the energy densities for DE ($\rho_{DE}$) in both models exhibit an increasing nature with redshift $(z)$ (which corresponds to a decreasing nature over cosmic time $(t)$) and remain positive throughout the expansion. 
For both models, $\rho_{DE}$ exhibits the expected positive behavior, indicating a contribution to the expansion of the universe. 
While, pressure of DE$(p_{DE})$ displays a negative behavior in both the present and the future. The universe may be expanding more quickly in this late era due to the pressure's negative nature. These observations are consistent with the accelerating universe's expanding nature.
\par In cosmological models, the Equation of State (EoS) parameter plays a crucial role in characterizing the nature of dark energy. The dependence of the energy density and pressure on each other is shown by the EoS parameter. Mathematically, the EoS parameter is given as $\omega = \frac{p}{\rho}$. By characterizing the nature of cosmic acceleration, the EoS parameter allows us to classify the universe's evolution into distinct regimes: The dust phase $(\omega = 0)$, the radiation-dominated phase $(\omega = \frac{1}{3})$, quintessence (\( -1 < \omega < -\frac{1}{3} \)), phantom (\( \omega < -1 \)) and cosmological constant era (\( \omega = -1 \)).
\vspace{0.3cm}\\
Using equation \eqref{equ(30)} and \eqref{equ(31)}, we obtain  $\omega_{DE}$ for model-1
\small
\begin{equation*}
    \omega_{DE} = \dfrac{ \alpha(1-2n)6^{n-1}\left[H_0 (z +1)^a (z+N)^{-b} N^ {\left(\frac{Nq_1}{N-1}\right)}  \right]^{2n} \left[3 - 2n(z +1) \left(\frac{a}{z+1} - \frac{b}{z+N} + \frac{q_1}{(z+1)^2} \ln (z+N) \right) \right]}{  \alpha \left(n - \frac{1}{2}\right) 6^n \left[H_0 (z +1)^a (z+N)^{-b} N^ {\left(\frac{Nq_1}{N-1}\right)}  \right]^{2n}  } \ \ \ \text{(for\  Model-1)}
    \tag{34} \label{equ(34}
\end{equation*}
\normalsize
where $a = 1 + q_0 -q_1 \ln N + \frac{q_1}{N-1}$ and $b = \frac{q_1}{z+1} + \frac{q_1}{N-1}$.
\vspace{0.3cm}\\
further, from eq. \eqref{equ(32)} and \eqref{equ(33)}, we get  $\omega_{DE}$ for model-2
\small
\begin{equation*}
    \omega_{DE} = \frac{-\alpha 6^{2n-1} \Bigg[3H_0^{2n}(1+z)^{3n} \exp\left(2dn \right)+2n H_0^{2n+1}(1+z)^{3n + \tfrac{5}{2}}\exp\left((2n+1) d \right) \left(\tfrac{3}{2} + \frac{zq_1 + q_2}{(1+z)^2} \right) \Bigg]}{\alpha \left(n- \frac{1}{2} \right)6^n \left[H_0 (1+z)^{\frac{3}{2}} \exp \left[d \right] \right]^{2n}} \ \ \ \text{(for\  Model-2)},
    \tag{35} \label{equ(35}
\end{equation*}
\normalsize
where $d = \frac{(q_1 + q_2)z^2 + 2 q_2 z}{2(1+z)^2}.$
\begin{figure}[h]
    \centering
     \includegraphics[width=9.0cm, height=6.0cm]{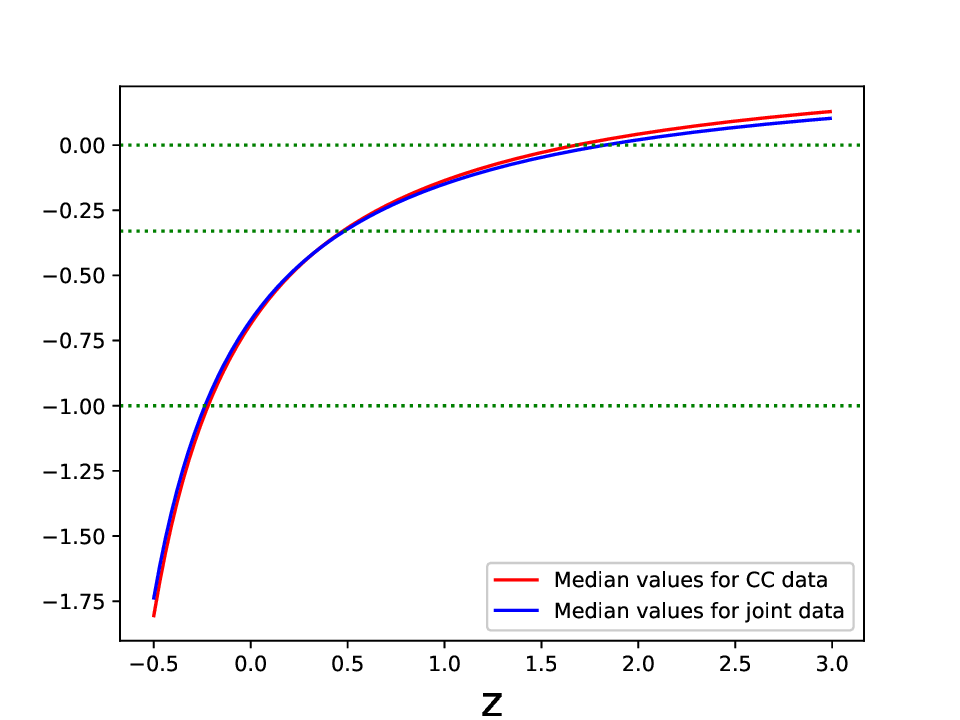}
    \caption{{\bf{For Model-1:}} $\omega_{DE}$ with $z$}
    \label{fig: Figure 6}
\end{figure}

\begin{figure}[h]
    \centering
     \includegraphics[width=9.0cm, height=6.0cm]{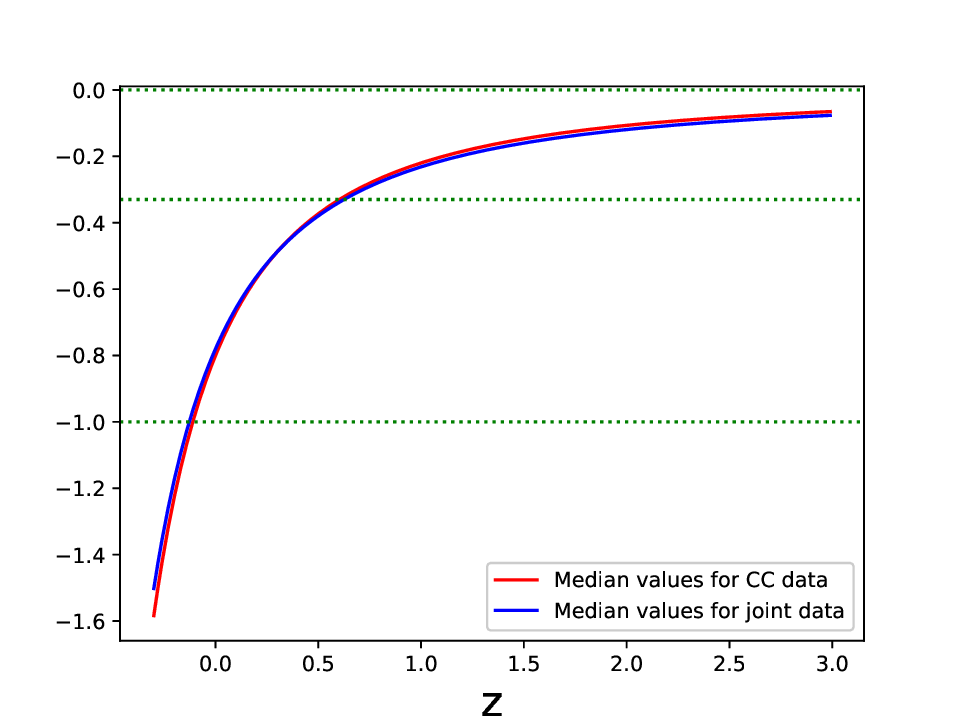}
    \caption{{\bf{For Model-2:}} $\omega_{DE}$ with $z$}
    \label{fig: Figure 11}
\end{figure} 
The graphical representation of $\omega_{DE}$ for both models' parameters is shown in figures (\ref{fig: Figure 6}) and (\ref{fig: Figure 11}).  For model-1, the value of the EoS parameter is $\omega_{DE} = -0.6857$ and $\omega_{DE} = -0.6722$ for CC data and Joint data, respectively, at $z =0$. Similarly, for model-2, the value of EoS parameter  are $\omega_{DE} = -0.8$(for CC data) and $\omega_{DE} = -0.78$ (for Joint data). 
The analysis of $\omega_{DE}$ shows that both models exhibit quintessence kind of dark energy for both data sets in the current epoch.  Moreover, for the median values of the model parameters, the universe eventually crosses the cosmological constant boundary $\omega = -1$ and exibit the phantom type dark energy in late time. We take $\alpha = 1$ and $n=1$ for these graphical illustrations. 


\subsection{Energy Conditions}
At a specific point in spacetime, the point-wise energy conditions that rely solely on the stress energy tensor are as follows \cite{singh2022lagrangian,GARG2025102025,visser1997energy}:
\begin{enumerate}
    \item Null Energy Condition (NEC): The sum of the energy density and pressure must always be greater than or equal to zero: $\rho_{eff} + p_{eff} \geq 0$.
    \item Weak Energy Condition (WEC): Both the energy density and the sum of the energy density and pressure must be non-negative: $  \rho_{eff} + p_{eff} \geq 0, \  \rho_{eff} \geq 0  $.
    \item Dominant Energy Condition (DEC): The energy density must be non-negative and $\rho_{eff} \geq |p_{eff}|$.
    \item Strong Energy Condition (SEC): This condition requires that: $\rho_{eff} +p_{eff} \geq 0$ and $\rho_{eff} + 3p_{eff} \geq 0$.
\end{enumerate}
\begin{figure}[h]
    \centering
    \includegraphics[width=9.0cm, height=6.4cm]{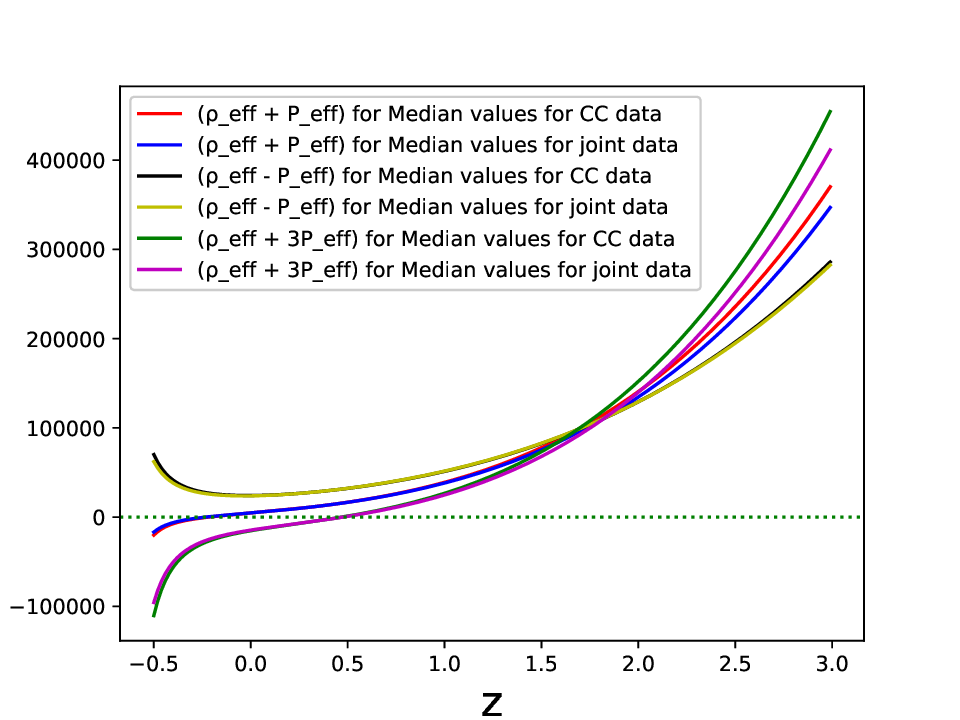}
    \caption{ {\bf{For Model-1:}} Components of energy conditions}
    \label{fig: Figure 14}
\end{figure}

\begin{figure}[h]
    \centering
     \includegraphics[width=9.0cm, height=6.4cm]{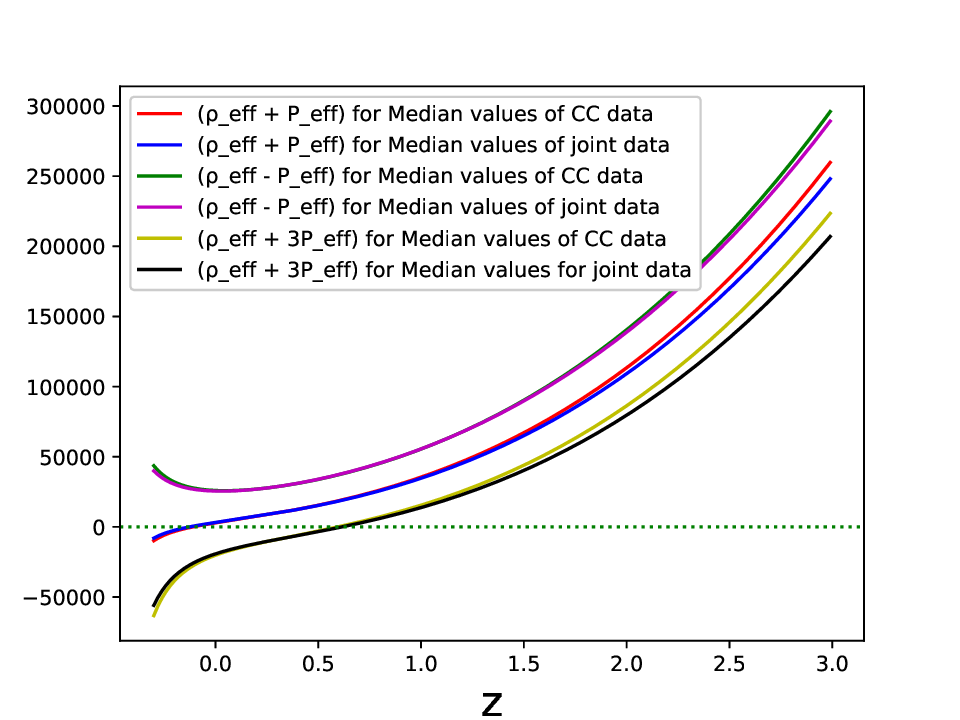}
    \caption{ {\bf{For Model-2}} Components of energy conditions}
    \label{fig: Figure 12}
\end{figure} 
The SEC  consists of the inequality $\rho_{eff} + 3p_{eff} \geq 0$, which may be connected to the Raychaudhuri equation for determining whether the universe's expansion is accelerating or decelerating\cite{singh2023homogeneous}. During a decelerating phase, the active gravitational mass ($\rho_{eff} + 3p_{eff}$) remains positive. However, observational data indicate that this condition is violated at some time between the epoch of galaxy formation and the present. Hence, the expansion of the universe at an accelerating rate (and consequently the presence of repulsive gravity) may be experienced for $\rho_{eff} + 3p_{eff} < 0$. This violation indicates the possibility that components with negative pressure exist, which exhibit anti-gravitational properties. Meanwhile, the Strong Energy Condition (SEC) contains two inequalities. It is important to mention that the violation of either one will violate the SEC\cite{singh2022lagrangian,  GARG2025102025, singh2023homogeneous}. \\[3pt]
Figures (\ref{fig: Figure 14}) and (\ref{fig: Figure 12}) graphically present all energy conditions for model-1 and model-2, respectively.  The plots show that both models satisfy the Null Energy Condition (NEC), Weak Energy Condition (WEC), and Dominant Energy Condition (DEC) up to the present time. Because the reconstructed model shows current accelerated expansion, the SEC (specifically $\rho_{eff} + 3 p_{eff} > 0$) is violated. The NEC is violated as the universe enters the phantom era, which may also lead to the violation of the WEC and DEC. The violation ($\rho_{eff} + p_{eff} \geq 0$)  indicates the presence of phantom-type dark energy in both models. Therefore, we can conclude that phantom dark energy may not be ruled out in both models. The violation of $\rho_{eff} + 3p_{eff} > 0$ will also be valid in the phantom dark energy dominated era. 


%% file: Chapters/STATEFINDERDIAGNOSTIC.tex
\subsection{Statefinder Diagnostic}
\begin{figure}[h]
    \centering
     \includegraphics[width=7.5cm, height=7.4cm]{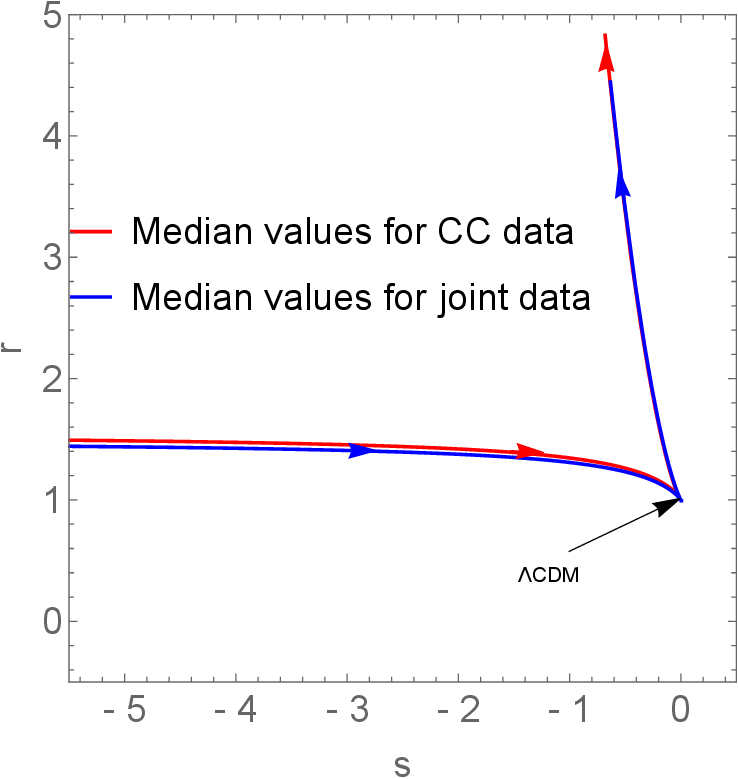}
    \caption{{\bf{For Model-1:}} s and r plane}
    \label{fig: Figure 7}
\end{figure}

\begin{figure}[h]
    \centering
     \includegraphics[width=7.5cm, height=7.4cm]{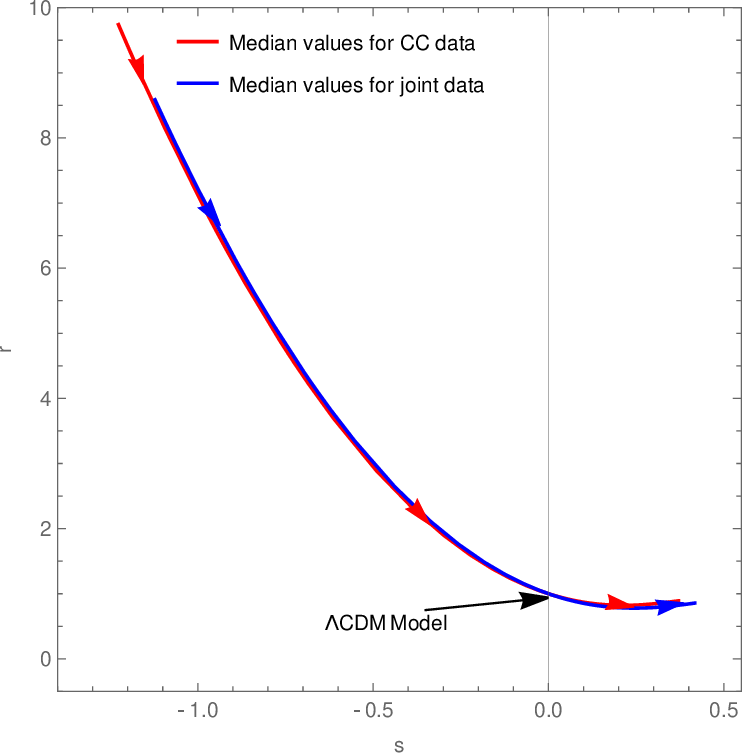}
    \caption{{\bf{For Model-2:}} s and r plane}
    \label{fig: Figure 13}
\end{figure}

It is well understood that geometric parameters can provide insights into the cosmological dynamics of a model. To explore alternative dark energy models beyond \(\Lambda\)CDM, it is crucial to examine additional parameters beyond the Hubble parameter \( H \) and the deceleration parameter \( q \). In this context, higher-order derivatives of the scale factor \( a(t) \), beyond \( H \) and \( q \), can serve as key elements in characterizing the dynamical nature of the universe.\\
The statefinder diagnostic, defined by geometric parameters $(r,s)$ \cite{38}, offers a useful way to analyze and differentiate dark-energy models by examining their evolution. The statefinder parameters $(r,s)$ are defined as:
\begin{equation*}
    r = \dfrac{\dddot a}{a H^3}, ~ s = \dfrac{r-1}{3(q -0.5)} \quad \text{where} \quad q \neq 0.5
\end{equation*}
Various dark energy models discussed in the literature can be characterized by different values of the statefinder pair $(r, s)$. In the r-s phase space, the Chaplygin gas model corresponds to the region where $r > 1$ and $s < 0$. The $\Lambda$CDM model is represented by the fixed point $(r = 1, s = 0)$, while quintessence models represent the region where $r < 1$ and $s > 0$. Holographic dark energy model given by the fixed point $(r = 1, s = \frac{2}{3})$. Additionally, the standard cold dark matter model is located at $(r = 1, s = 1)$ in the r-s plane.
\vspace{0.3cm}\\
The  $(s, r)$ trajectories for Model-1 and Model-2 are illustrated in figures (\ref{fig: Figure 7}) and (\ref{fig: Figure 13}), respectively. For model-1, the figure (\ref{fig: Figure 7}) demonstrates the evolutionary trajectory of the statefinder pair $(s, r)$, which originates in the Chaplygin gas regime at early time, passes through the $\Lambda$CDM  point, and evolves toward a unified description incorporating both dark matter and dark energy components. In figure (\ref{fig: Figure 13}), the trajectory of the $(s, r)$ parameters begins with Chaplygin gas models, passes through the $\Lambda$CDM point, and evolves toward quintessence models in the present epoch for both data estimates.

\subsection{Age of the Universe}
The cosmic age $t(z)$ of cosmological model can be expressed in terms of redshift $z$ as\cite{40} 
\begin{equation*}
    t(z) = \int_z^{\infty} \dfrac{dz}{(1+z) H(z)}
\end{equation*}
Using the Hubble parameters $H(z)$ (from equation \eqref{equ(21)} and \eqref{equ(23)}), we numerically evaluate this integral at the present epoch $(z=0)$ to determine the current age of the universe. For the model-1, the present age is $t_0$ (at $z=0) = 13.07$ Gyr and $t_0= 13.19$ Gyr for the CC and joint dataset, respectively. For the model-2, the present age is $t_0$ $=13.67$ Gyr (CC) and $t_0 = 13.78$ Gyr (joint), both of which are close to the current age values of the $\Lambda$CDM model (having $t_0 = 13.79$ Gyr) obtained from Planck results\cite{2020A&A...641A...6P}.

%% file: Chapters/CONCLUSION.tex
\section{Conclusion}\label{sec:6}
The behavior of the universe for a spatially flat FLRW metric has been examined in the context of $f(T)$ gravity theory, considering power law functional form, $ f(T) = \alpha (-T)^n.$ Specifically, we adopted two distinct parameterization forms of the deceleration parameter which are given by \(q(z) = q_0 + q_1 \left( \frac{\ln(N +z)}{z+1} - \ln N \right)\) and $q(z) = \dfrac{1}{2} + \dfrac{q_1 z + q_2}{(1+z)^2}$. We determine the median values of the model parameters by  using the Bayesian analysis MCMC analysis in emcee tool. The parameters are constrained through the MCMC analysis for both models and the results are summarized in table (\ref{tab: Table 1}) and table (\ref{tab: Table 2}). The best-fit curve of $H(z)$ illustrates the compatibility of the both models with the measured cosmic chronometer observations.

\par We further examined the evolutionary dynamics using various cosmological parameters. Furthermore, the analysis includes a discussion on the behavior of the deceleration parameter. The evolution curve of the deceleration parameter (figures (\ref{fig: Figure 3}) and (\ref{fig: Figure 8})) illustrate the transition of the Universe from decelerated expansion to accelerated expansion. For Model-1, the current value of deceleration parameter is $q_0 = -0.648$ (CC) and $q_0 = -0.622$ (joint), the value of transition redshift is $z_t = 0.62$ (CC) and $z_t = 0.63$(joint). For Model-2, the present value of deceleration parameter is obtain $q_0 = -0.7$ and $ -0.67$ for CC and joint data respectively. The value of transition redshift is $z_t= 0.60$ (CC) and $z_t=0.62$(joint). 
The negative sign of $q_0$ confirms that cosmic expansion is accelerating  at the current epoch $(z =0)$ for both models. These results align with the expanding nature of the accelerating Universe. 

\par Additionally, we analyzed the behavior of physical parameters. The energy density for DE ($\rho_{DE}$) remains positive throughout cosmic evolution while pressure for DE ($p_{DE}$) becomes negative in the recent due to dominance of dark energy.  The model demonstrates that the observed cosmic acceleration arises from the presence of negative pressure during the late-time era of universal expansion.  
Based on the constrained parameter values, the behavior of the EoS parameter($\omega_{DE}$) for DE  is depicted in the figures  (\ref{fig: Figure 6}) and (\ref{fig: Figure 11}). For model-1, the values of the EoS parameter are $\omega_{DE} = -0.6857$ and $\omega_{DE} = -0.6722$ at $z=0$ for the CC and joint estimates, respectively. Similarly, For model-2, the values of the EoS parameter is $\omega_{DE}=-0.8$ (CC) and $\omega_{DE}=-0.78$ (joint) at present. The analysis of $\omega_{DE}$ shows that both models exhibit quintessence kind of dark energy in the current epoch. The EoS parameter currently lies in the quintessence region but exhibits the phantom region in the future. 

\par Lastly, we obtained the present age of the universe for both models. For Model-1, the age of the universe is $t_0$ = $13.07$ Gyr and $t_0= 13.19$ Gyr for CC data and for the joint dataset, respectively, and for Model-2, the age of the universe is $t_0$ = $13.67$ Gyr(CC) and $t_0= 13.78$ Gyr (joint). In conclusion, our overall observation is that both models demonstrate consistency with observational evidence for the current epoch of accelerated cosmic expansion.
\section*{Acknowledgements}
 G. P. Singh is thankful to the Inter-University Centre for Astronomy and Astrophysics (IUCAA), Pune, India, for support under the Visiting Associateship programme. Gauree Shanker is thankful to the Department of Science and Technology (DST), Government of India, for providing financial assistance in terms of the FIST project (TPN-69301) vide the letter with Ref. No.: (SR/FST/MS-1/2021/104).
